 \documentclass[rnote]{aa} 
\usepackage{graphicx}
\usepackage{amssymb}
\usepackage{txfonts}
\begin{document}

   \title{Reanalysis of the FEROS observations of HIP~11952\thanks{
   Based on data products from observations made with ESO Telescopes at the La Silla Paranal Observatory under programme ID 60.A-9036, 072.C-0488, 072.C-0513, 073.C-0784, 074.C-0012, 074.D-0380, 075.C-0234, 075.D-0760, 076.C-0073, 076.C-0878, 077.A-9009, 077.C-0138, 077.C-0192, 077.C-0530, 078.A-9048, 078.C-0378, 078.C-0833, 079.A-9006, 079.A-9017, 079.C-0170, 079.C-0681, 080.A-9005, 080.A-9021, 080.C-0032, 082.A-9011, 082.C-0315, 083.A-9011, 084.A-9003, 084.A-9003, 084.A-9004, 084.A-9004, 084.A-9011, 085.A-9027, 085.A-9027, 085.C-0557, 086.A-9006, 086.A-9006, 086.A-9014, 086.A-9014, 086.D-0460, 087.A-9014, 087.C-0476, 088.A-9007, 088.A-9007.}}

   \subtitle{}

   \author{
A. M\"{u}ller\inst{1},V.~Roccatagliata\inst{2}, Th. Henning\inst{3}, 
D. Fedele\inst{4},
A. Pasquali\inst{5}, 
E. Caffau\inst{6}, 
M. V. Rodr{\'i}guez-Ledesma\inst{3,7}, M. Mohler-Fischer\inst{3},
U. Seemann\inst{7} \and R. J. Klement\inst{8,3}
         }

   \institute{European Southern Observatory, Alonso de Cordova 3107, Vitacura, Santiago, Chile\\\email{amueller@eso.org} \and
   Universit\"ats-Sternwarte M\"unchen, Ludwig-Maximilians-Universit\"at, Scheinerstr.~1, 81679 M\"unchen, Germany \and
  Max-Planck-Institut f\"ur Astronomie, K\"onigstuhl 17, 69117 Heidelberg, Germany \and
  Max Planck Institut f\"ur Extraterrestrische Physik, Giessenbachstrasse 1, 85748 Garching, Germany \and
  Astronomisches Rechen-Institut, Zentrum f\"ur Astronomie, M\"onchhofstrasse, 12-14, 69120 Heidelberg, Germany \and
  Zentrum f\"ur Astronomie der Universit\"at Heidelberg (ZAH), Landessternwarte, K\"onigstuhl 12, 69120 Heidelberg, Germany \and
  Institut f\"ur Astrophysik, Georg-August-Universit\"at, Friedrich-Hund-Platz 1, 37077 G\"ottingen, Germany \and
  Department of Radiotherapy and Radiation Oncology, Leopoldina Hospital, Gustav-Adolf-Str. 8, 97422 Schweinfurt, Germany
             }

   \date{Received XX March 2013; Accepted 2013}

 
  \abstract
   {}
   {We reanalyze FEROS observations of the star HIP~11952 to reassess the existence of the proposed planetary system.}
   {The radial velocity of the spectra were measured by cross-correlating the observed spectrum with a synthetic template. We also analyzed a large dataset of FEROS and HARPS archival data of the calibrator HD~10700 spanning over more than five years. We compared the barycentric velocities computed by the FEROS and HARPS pipelines.}
  {The barycentric correction of the FEROS-DRS pipeline was found to be inaccurate and to introduce an artificial one-year period with a semi-amplitude of 62~m\,s$^{-1}$. Thus the reanalysis of the FEROS data does not support the existence of planets around HIP~11952.}
   {}

   \keywords{techniques: spectroscopic - stars: planetary systems - techniques: radial velocities - stars: individual: HIP 11952
                }

   \titlerunning{FEROS observations of HIP~11952} 
   \authorrunning{A. M\"{u}ller et al.}
  \maketitle
%

\section{Introduction}

In \citet{Setiawanetal2012} we reported the discovery of two planets around \object{HIP~11952}, a metal-poor 
star with $[Fe/H]=-1.9$~dex. 
We reported a long-period radial velocity (RV hereafter) variation of 290~d and a short-period 
variation of 6.95~d based on FEROS observations with the ESO/MPG 2.2 m telescope in La Silla.
The spectroscopic analysis of the stellar
activity revealed a stellar rotation period of 4.8~d. The observed RV variations had 
been associated to two planets with minimum planetary masses of 0.78~M$_{\rm Jup}$ 
and 2.93~M$_{\rm Jup}$ for the inner and the outer planet, respectively. 
In contrast to this result, based on HARPS N (mounted at the Telescopio Nazionale Galileo, La Palma) and HARPS (mounted at the ESO 3.6-m telescope in La Silla, Chile) observations, \citet[][]{Desideraetal2013} did not find any evidence for planets around HIP~11952. In this research note we reanalyze the RV signal of the FEROS data to clarify this discrepancy. In a first step we used the star HD~10700 to compare FEROS and HARPS 
measurements since there are many observations available for this object taken with both instruments. 
Recently, \citet{Tuomietal2012} found a signal of five significant periodicities in the RV of 
HD~10700 that might be caused by a system of five planets with minimum masses 
between 2.0 and 6.6~M$_\oplus$. Since those signals are associated to RV variations below 1~m\,s$^{-1}$, 
the object is still suited to compare the FEROS \citep[][]{Kauferetal1999} and HARPS \citep[][]{Mayoretal2003} data, 
taking into account the long-term precision of the FEROS RV measurements of about $>$10~m\,s$^{-1}$. 
The comparison reveals that the barycentric correction of the FEROS pipeline data is less acccurate than needed 
for long-period RV monitoring. We applied a much more accurate correction to the FEROS data of HIP~11952 and 
show that with this improved correction there is indeed no evidence for planets around this object.
\\\\
The paper is organized as follows: Sections~\ref{secrvhd10700} and \ref{secrvhip} present the observations and RV measurements of HD~10700 and HIP~11952, respectively. The effect of the inaccurate barycentric correction on FEROS RV measurements is presented in Sec.~\ref{seceffect}. The conclusions can be found in Sec.~\ref{seccon}.


\section{Radial velocity measurements of HD~10700}\label{secrvhd10700}
\begin{figure}
\centering
\includegraphics[width=9cm]{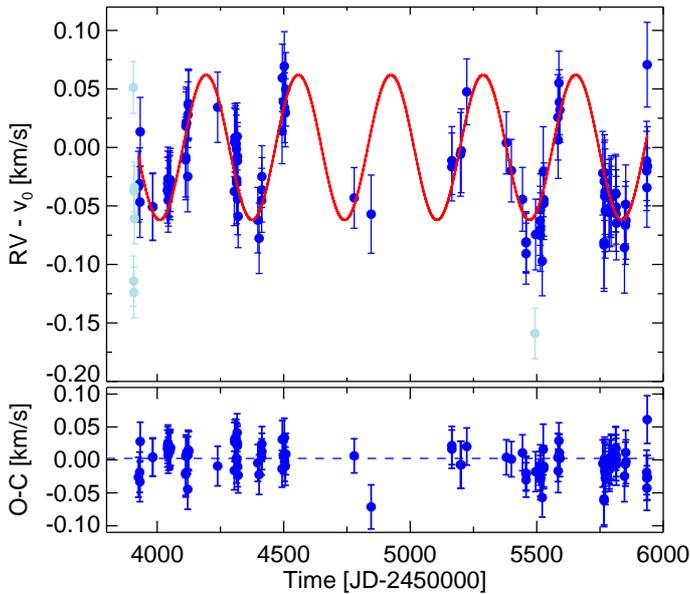}
\caption{RV measurements of HD~10700 using spectra reduced by FEROS-DRS. 
The red solid line is a sinusoidal fit with a period of 365.2$\pm$0.2~days and a semi-amplitude
of 62.0$\pm$1.1~m\,s$^{-1}$. Only spectra with a simultaneous ThAr exposure (blue data points) were considered for the fit. The present period of 365.2 days is caused by an inaccurate barycentric correction by the FEROS-DRS. The lower panel shows the residual of the fit. }
\label{hd10700_feros}
\end{figure}

\begin{figure}
\centering
\includegraphics[width=9cm]{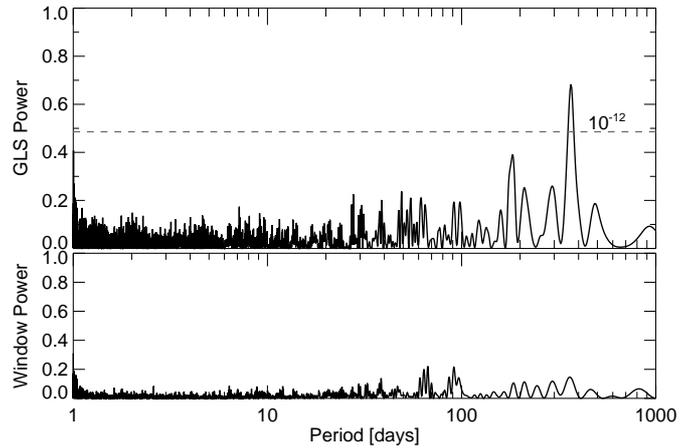}
\caption{Generalized Lomb-Scargle periodogram of the RV values of HD~10700. The horizontal dashed line indicates an FAP = 10$^{-12}$. A significant period of 365.2 days with an FAP of $7\cdot10^{-24}$ is present in the FEROS 
data set. }
\label{hd10700_ferosgls}
\end{figure}

We analyzed a large data set of FEROS (117) and HARPS (153) spectra of the  
star \object{HD~10700}. The considered FEROS observations span a period of 5.6~years, while the 
HARPS observations cover 5.3~years. The measurement of the stellar RVs from FEROS and HARPS spectra was carried out using a self-written routine that cross-correlates the stellar spectra with a template spectrum. The template was a synthetic spectrum representing the stellar parameters of HD~10700. We computed of the synthetic spectrum using {\sc spectrum} \citep{gra94} with the ATLAS9 atmosphere models \citep{cas04}. Figure~\ref{hd10700_feros} shows the RV measurements
based on FEROS data with the barycentric correction originally applied by the FEROS-DRS pipeline. 
Only exposures with a simultaneous ThAr spectrum were used to perform a sinusoidal fit.
The radial velocity reveals a period  of 365.2~days with a semi-amplitude of 62~m\,s$^{-1}$.
Figure~\ref{hd10700_ferosgls} shows the computed generalized Lomb-Scargle (GLS) periodogram \citep{ZechmeisterKurster2009}. The one-year period is clearly present and visible as the highest peak in the GLS. It has a false-alarm probability (FAP hereafter) of $7\cdot10^{-24}$ computed by the GLS.

\begin{figure}
\centering
\includegraphics[width=9cm]{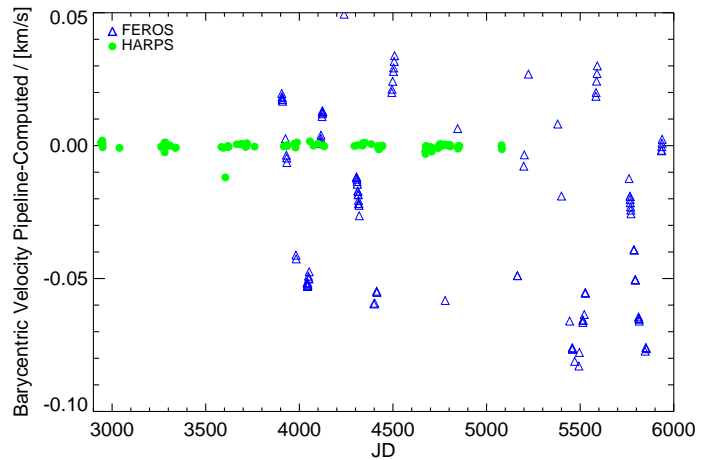}
\caption{Difference of the computed and pipeline-provided barycentric correction for HD~10700. Green circles represent HARPS data and blue triangles represent FEROS data. }
\label{HD10700_feros_harps}
\end{figure}

We found that the long period in the FEROS data arises because 
the barycentric correction of the FEROS-DRS pipeline was not accurate enough.
In particular, it neglects the precession of the coordinates. 
We applied a more precise barycentric correction using the IDL code {\tt baryvel.pro}\footnote[1]{Part of the IDL Astronomy Users Library available at\\ \url{http://idlastro.gsfc.nasa.gov}}. The algorithm is based on \citet{Stumpff1980} and is accurate to $\sim$1~m\,s$^{-1}$.

In Figure~\ref{HD10700_feros_harps} we show the difference of the IDL-computed and pipeline-provided barycentric 
correction for HD~10700.  

\begin{figure}
\centering
\includegraphics[width=9cm]{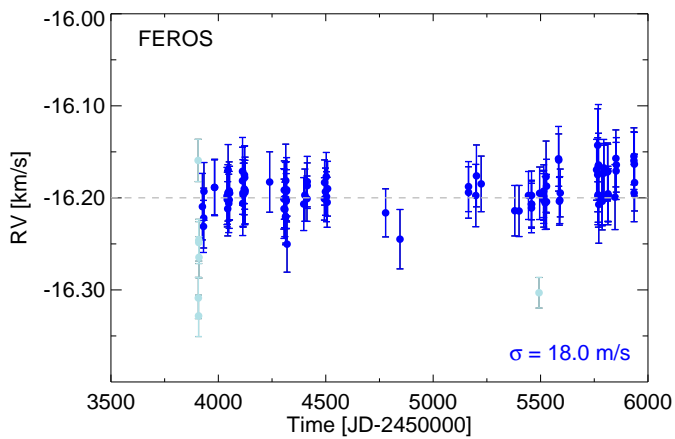}
\caption{RV measurements of HD~10700 using FEROS data. An accurate barycentric correction was applied.  
Light-blue data points indicate an observing mode where no simultaneous ThAr spectrum was taken.}
\label{new_feros_hd10700}
\end{figure}

\begin{figure}
\centering
\includegraphics[width=9cm]{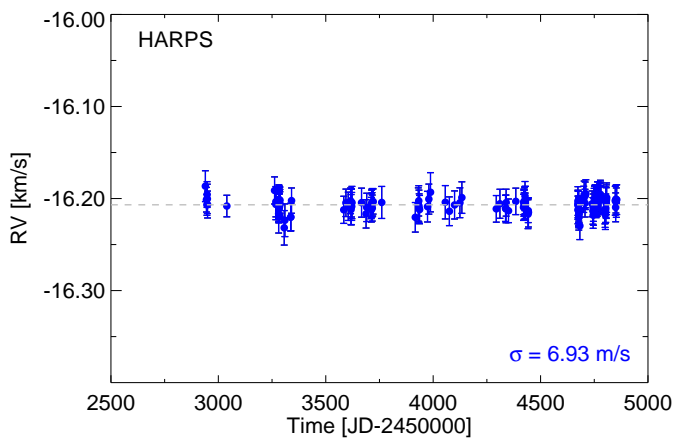}
\caption{RV measurements of HD~10700 using HARPS data.}
\label{new_harps_hd10700}
\end{figure}

The comparison between the values computed by baryvel.pro and by the HARPS pipeline (in green) demonstrates that the adopted 
procedure computes the barycentric correction with the required accuracy, in contrast to the FEROS-DRS pipeline (in blue). The RMS-dispersion of the difference of the barycentric correction is 1.2~m\,s$^{-1}$ and 34.6~m\,s$^{-1}$ for HARPS-baryvel.pro and FEROS-baryvel.pro, respectively.
Applying an accurate barycentric correction to the FEROS spectra, the RV values of HD~10700 are 
basically constant (see Figure~\ref{new_feros_hd10700}). No indication of either a one-year period or any other period is left. 
We also applied this routine to the HARPS data of HD~10700. The result is shown in Figure~\ref{new_harps_hd10700}. 
In addition to the higher signal-to-noise ratio (S/N) and spectral resolution of HARPS data with respect to the FEROS data, it should be considered that HARPS is an instrument fully 
optimized for high radial velocity precision (e.g. improved stability, vacuum tank).

\section{Radial-velocity measurements of HIP~11952}\label{secrvhip}
\subsection{Observations}

The dataset of HIP~11952 includes the 77 observations presented in \citet{Setiawanetal2012} carried out 
from August 2009 until January 2011. 
We also included seven new FEROS observations not present in the analysis of \citet{Setiawanetal2012}.
All details about the observations are reported in Table~\ref{ob}, including the S/N
values measured at 5520\AA.

We noticed that the data obtained between October 22 and 25, 2010 had poor ThAr+Ne frames.

\subsection{Analysis}
\begin{figure}
\centering
\includegraphics[width=9cm]{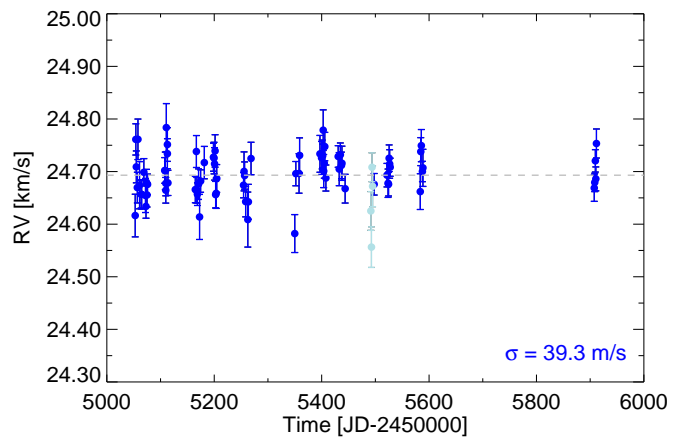}
\caption{RV measurements of the full data set of HIP~11952 containing 84 observations. The synthetic template
was used. Light-blue data points indicate an observing mode where no simultaneous ThAr spectrum was taken.
The standard deviation of the data set is 39~m\,s$^{-1}$, which resembles the size of the error bars.}
\label{rv_hip11952}
\end{figure}

The RV of HIP~11952 was extracted by cross-correlating
the stellar spectrum using two different templates and applying the newly computed barycentric correction. First, we used a synthetic spectrum created using the {\it Synthe} code based on ATLAS models with the stellar parameters determined in \citet{Setiawanetal2012} (e.g. $T_{eff}$=5960~K, 
$\log g$=3.8 ($g$ in~cm\,s$^{-2}$), $[Fe/H]=-1.95$~dex). As a second approach, we used the stellar spectrum of HIP~11952 itself. We chose the observation with the highest
S/N (S/N=170) of all observations obtained on January 3, 2010.
In both cases the RV was extracted in each FEROS order. 
The final RV was obtained using 15 echelle orders, ranging between 4143\AA~and 6402\AA~in 
wavelength. These orders have the highest S/N and are free of telluric lines. The RV values together with their errors are reported in Table~\ref{ob}
for the synthetic template. We obtained similar results with the observed spectrum of
HIP~11952. We obtained standard deviations of 36.1~m\,s$^{-1}$ using the stellar template,
and 39.3~m\,s$^{-1}$ using the synthetic template (see Figure~\ref{rv_hip11952}).

\begin{figure}
\centering
\includegraphics[width=9cm]{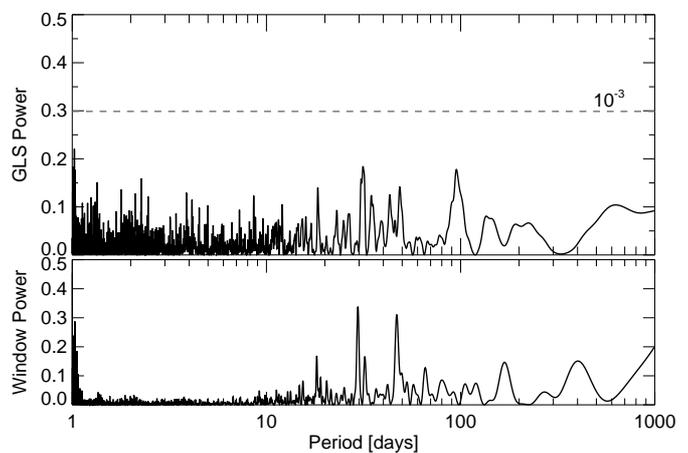}
\caption{GLS periodogram of the RV values of HIP~11952. There is no period present in the FEROS data set. 
The horizontal dashed line indicates an FAP = 10$^{-3}$.}
\label{GLS_hip11952}
\end{figure}

We applied the GLS periodogram to the RV data to search for periodicities (see Figure~\ref{GLS_hip11952}).
We found no significant period in the FEROS data. In particular, we found no sign of the 290\,d and 6.95\,d periods claimed 
in \citet{Setiawanetal2012}, in agreement with the HARPS~N result presented by \citet{Desideraetal2013}. 

\section{Effect of the inaccurate barycentric correction on RV measurements}\label{seceffect}
The inaccuracy of the barycentric correction of the pipeline might also affect previous RV studies carried out with FEROS. To investigate the effect on period searches using FEROS RV measurements without the applied accurate barycentric correction, we carried out a simulation using the following setup. As time basis we adopted the 77 epochs published in \citet{Setiawanetal2012}, which cover 536~days. For periods between 5 and 350~days and velocity semi-amplitudes between 25 and 250~m\,s$^{-1}$, sine functions with noise were computed. Error bars with a mean value of $\sim$17~m\,s$^{-1}$ were added. The RMS-dispersion of the difference for the noisy sine functions minus the sinus function without noise is $\sim$17~m\,s$^{-1}$, too. The sinusoidal fit displayed in Fig.~\ref{hd10700_feros} was added to each computed sine function to account for the inaccurate barycentric correction by the FEROS-DRS. For every setup we computed a GLS periodogram to determine whether the applied period could still be identified. 
Figure~\ref{simgls} shows the computed GLS periodogram for each setup. The period and velocity semi-amplitude values used to 
compute the individual sine functions are displayed in the legend of the individual plots. The red dashed line represents the position of the period applied to compute the individual sine functions. 
\newline
For pure sinusoidal RV variations, i.e. for variations without eccentricity, we draw the following conclusions:
\begin{itemize}
 \item All periods can be identified for RV variations with semi-amplitudes greater than $\sim$150~m\,s$^{-1}$.
 \item For semi-amplitudes of around 100~m\,s$^{-1}$ periods between $\sim$100 and $\sim$200~days cannot be clearly identified.
 \item If the semi-amplitude is smaller than 100~m\,s$^{-1}$, all periods present in the RV data set cannot be identified or are difficult to identify.
 \item Short periods $\lesssim$10~days can be detected reliably if the RV amplitude is greater than about twice the RV residuals.
\end{itemize}

%

\section{Conclusions}\label{seccon}

In this research note we presented the analysis of the HARPS and FEROS spectra of the star 
HD~10700 and a new detailed re-analysis of the FEROS spectra of HIP~11952, in which we included seven new observations not presented in \citet{Setiawanetal2012}.
We found that
   \begin{enumerate}
      \item the barycentric correction of the FEROS pipeline was not accurate enough based on the analysis of HD~10700, and
      \item the RV of HIP~11952 was recomputed applying a proper barycentric 
      correction. As a result, no significant periodicity was found in HIP~11952.
   \end{enumerate}
This investigation clarifies the dsicrepancy between the FEROS-based RV measurements and the HARPS-based
measurements and confirms the conlcusion that we see no evidence for planets around HIP~11952.

\begin{acknowledgements}
V.R. was supported by the DLR grant number 50~OR~1109 and by the {\it Bayerischen Gleichstellungsf{\"o}rderung} (BGF). 
\end{acknowledgements}
     
\bibliographystyle{aa}
\bibliography{references}

\begin{thebibliography}{9}
\expandafter\ifx\csname natexlab\endcsname\relax\def\natexlab#1{#1}\fi

\bibitem[{{Castelli} \& {Kurucz}(2004)}]{cas04}
{Castelli}, F. \& {Kurucz}, R.~L. 2004, ArXiv Astrophysics e-prints

\bibitem[{Desidera {et~al.}(2013)Desidera, Sozzetti, Bonomo, Gratton, Poretti,
  Claudi, Latham, Affer, Cosentino, Damasso, Esposito, Giacobbe, Malavolta,
  Nascimbeni, Piotto, Rainer, Scardia, Schmid, Lanza, Micela, Pagano, Bedin,
  Biazzo, Borsa, Carolo, Covino, Faedi, Hebrard, Lovis, Maggio, Mancini,
  Marzari, Messina, Molinari, Munari, Pepe, Santos, Scandariato, Shkolnik, \&
  Southworth}]{Desideraetal2013}
Desidera, S., Sozzetti, A., Bonomo, A., {et~al.} 2013, A\&A, in press

\bibitem[{{Gray} \& {Corbally}(1994)}]{gra94}
{Gray}, R.~O. \& {Corbally}, C.~J. 1994, AJ, 107, 742

\bibitem[{{Kaufer} {et~al.}(1999){Kaufer}, {Stahl}, {Tubbesing},
  {N{\o}rregaard}, {Avila}, {Francois}, {Pasquini}, \&
  {Pizzella}}]{Kauferetal1999}
{Kaufer}, A., {Stahl}, O., {Tubbesing}, S., {et~al.} 1999, The Messenger, 95, 8

\bibitem[{{Mayor} {et~al.}(2003){Mayor}, {Pepe}, {Queloz}, {Bouchy},
  {Rupprecht}, {Lo Curto}, {Avila}, {Benz}, {Bertaux}, {Bonfils}, {Dall},
  {Dekker}, {Delabre}, {Eckert}, {Fleury}, {Gilliotte}, {Gojak}, {Guzman},
  {Kohler}, {Lizon}, {Longinotti}, {Lovis}, {Megevand}, {Pasquini}, {Reyes},
  {Sivan}, {Sosnowska}, {Soto}, {Udry}, {van Kesteren}, {Weber}, \&
  {Weilenmann}}]{Mayoretal2003}
{Mayor}, M., {Pepe}, F., {Queloz}, D., {et~al.} 2003, The Messenger, 114, 20

\bibitem[{{Setiawan} {et~al.}(2012){Setiawan}, {Roccatagliata}, {Fedele},
  {Henning}, {Pasquali}, {Rodr{\'{\i}}guez-Ledesma}, {Caffau}, {Seemann}, \&
  {Klement}}]{Setiawanetal2012}
{Setiawan}, J., {Roccatagliata}, V., {Fedele}, D., {et~al.} 2012, \aap, 540,
  A141

\bibitem[{{Stumpff}(1980)}]{Stumpff1980}
{Stumpff}, P. 1980, \aaps, 41, 1

\bibitem[{{Tuomi} {et~al.}(2013){Tuomi}, {Jones}, {Jenkins}, {Tinney},
  {Butler}, {Vogt}, {Barnes}, {Wittenmyer}, {O'Toole}, {Horner}, {Bailey},
  {Carter}, {Wright}, {Salter}, \& {Pinfield}}]{Tuomietal2012}
{Tuomi}, M., {Jones}, H.~R.~A., {Jenkins}, J.~S., {et~al.} 2013, \aap, 551, A79

\bibitem[{{Zechmeister} \& {K{\"u}rster}(2009)}]{ZechmeisterKurster2009}
{Zechmeister}, M. \& {K{\"u}rster}, M. 2009, \aap, 496, 577

\end{thebibliography}


\appendix

\section{GLS periodograms for simulated RV measurements}

\begin{figure}
\onecolumn
\centering
\includegraphics[angle=90, width=17cm]{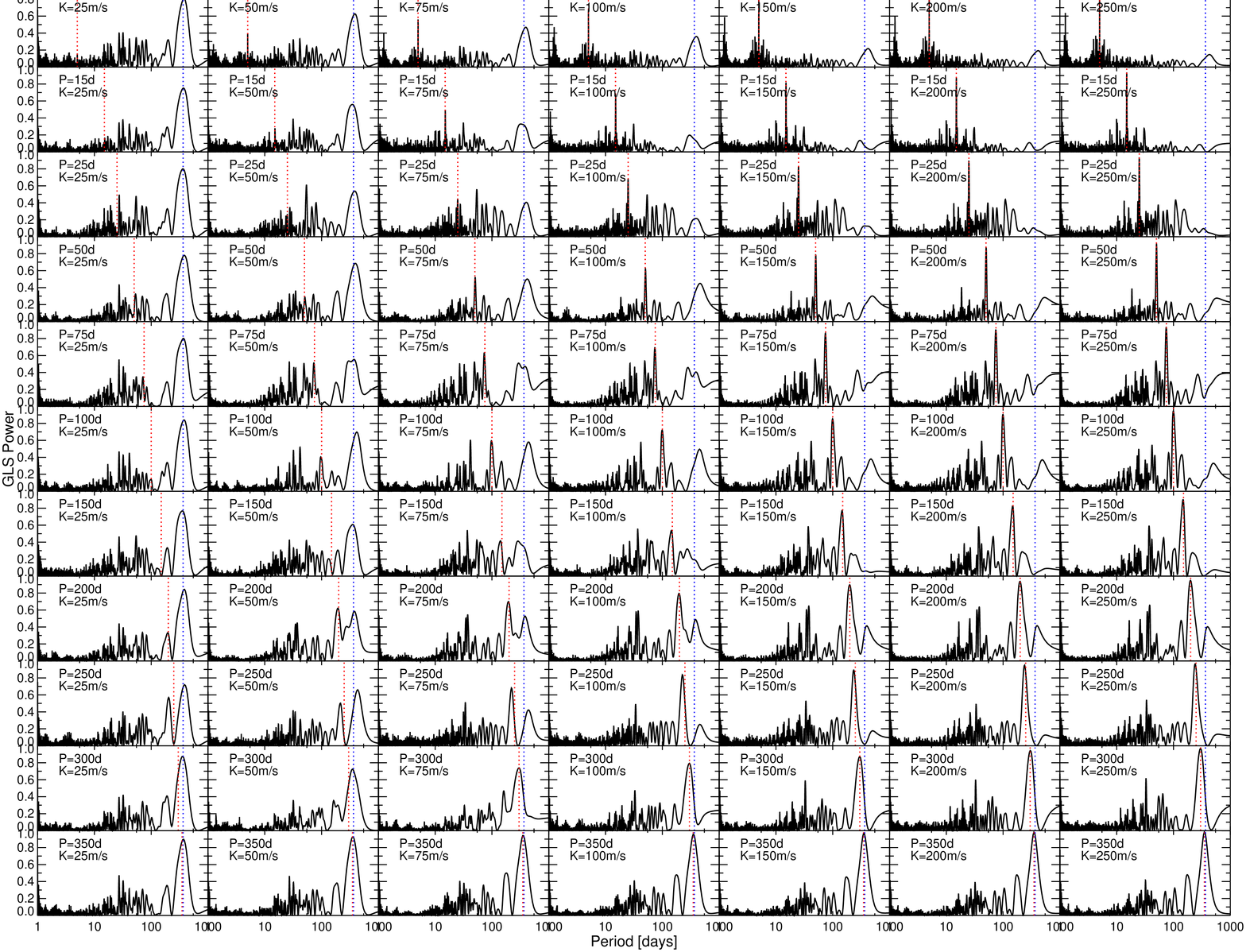}
\caption{GLS periodogram for sine functions with different periods and velocity semi-amplitudes. The fit displayed in Fig.~\ref{hd10700_feros} was added to each sine function to account for the inaccurate barycentric correction by the FEROS-DRS. The period and velocity semi-amplitude used to compute the sine function are displayed in the legend of the individual plots. The red dashed line represents the position of the period applied to compute the sine function. The blue dashed line marks a one-year period.}
\label{simgls}
\end{figure}
\twocolumn

\section{Accompanying table}

\onecolumn
\begin{longtab}
\centering
\begin{longtable}{lcccccc}
\caption{\label{ob} Observation log.}\\ 
\hline\hline                
\noalign{\smallskip}
JD / [JD-240\,000 d]	&	Exp.Time / [second]	&	S/N	&	Airmass	&	Seeing / [arcsec]	& RV / [m\,s$^{-1}$]	&	RVerr / [m\,s$^{-1}$]	\\
\noalign{\smallskip}
\hline                
\noalign{\smallskip}
\hline
\endfirsthead
\caption{continued.}\\
\hline
\hline
\noalign{\smallskip}
JD / [JD-240\,000 d]	&	Exp.Time / [second]	&	S/N	&	Airmass	&	Seeing / [arcsec]	& RV / [m\,s$^{-1}$]	&	RVerr / [m\,s$^{-1}$]	\\
\noalign{\smallskip}
\hline
\endhead
\hline
\endfoot
\noalign{\smallskip}
55052.83202	&	600		&	59	&	1.192	&	1.61		&	24616.40	&	40.54	\\
55053.94928	&	600		&	62	&	1.069	&	0.83		&	24761.19	&	29.71	\\
55054.90250	&	600		&	89	&	1.046	&	0.80		&	24708.95	&	30.31	\\
55056.78339	&	900		&	70	&	1.385	&	1.24		&	24669.57	&	29.18	\\
55057.94419	&	900		&	106	&	1.080	&	1.12		&	24761.33	&	38.50	\\
55060.87386	&	960		&	41	&	1.052	&	2.16		&	24674.07	&	44.89	\\
55063.83403	&	900		&	84	&	1.095	&	2.51		&	24656.70	&	26.10	\\
55064.88164	&	900		&	146	&	1.045	&	0.94		&	24656.71	&	27.92	\\
55068.89749	&	900		&	144	&	1.058	&	0.98		&	24698.63	&	25.98	\\
55071.92782	&	900		&	154	&	1.128	&	0.79		&	24654.55	&	32.41	\\
55072.90426	&	900		&	151	&	1.082	&	0.92		&	24633.79	&	22.48	\\
55073.76953	&	900		&	152	&	1.207	&	1.01		&	24679.67	&	24.56	\\
55074.91064	&	900		&	158	&	1.106	&	0.73		&	24655.54	&	22.92	\\
55075.82762	&	900		&	127	&	1.057	&	1.30		&	24675.65	&	19.60	\\
55107.78074	&	900		&	25	&	1.050	&	--		&	24702.16	&	24.39	\\
55108.84721	&	800		&	57	&	1.192	&	1.02		&	24701.29	&	35.77	\\
55109.76234	&	1200	&	118	&	1.046	&	1.07	&		24664.45	&	24.41	\\
55109.78002	&	800		&	136	&	1.052	&	1.07		&	24677.58	&	23.65	\\
55110.77245	&	800		&	53	&	1.050	&	0.75		&	24783.65	&	45.61	\\
55112.76892	&	900		&	141	&	1.051	&	~0.6		&	24734.04	&	28.53	\\
55112.78081	&	800		&	141	&	1.061	&	~0.5		&	24751.45	&	32.03	\\
55113.76268	&	900		&	164	&	1.049	&	0.60		&	24678.47	&	23.97	\\
55164.74363	&	900		&	150	&	1.495	&	0.63		&	24665.81	&	25.04	\\
55166.74047	&	1000	&	151	&	1.522	&	0.54	&		24738.16	&	30.37	\\
55167.63152	&	1000	&	169	&	1.063	&	0.66	&		24664.44	&	24.32	\\
55168.64106	&	900		&	161	&	1.080	&	0.65		&	24657.06	&	21.53	\\
55170.73747	&	1000	&	139	&	1.603	&	0.65	&		24676.07	&	29.07	\\
55172.68382	&	900		&	115	&	1.244	&	0.84		&	24613.74	&	42.86	\\
55174.66913	&	1200	&	170	&	1.211	&	1.21	&		24681.98	&	21.38	\\
55181.54032	&	1000	&	157	&	1.054	&	0.70	&		24717.05	&	30.76	\\
55198.52849	&	900		&	163	&	1.048	&	0.64		&	24727.36	&	26.69	\\
55199.60741	&	900		&	171	&	1.231	&	0.71		&	24725.67	&	30.05	\\
55200.62188	&	900		&	135	&	1.322	&	1.00		&	24714.64	&	30.84	\\
55201.58493	&	900		&	150	&	1.164	&	1.28		&	24739.26	&	30.82	\\
55202.61649	&	900		&	140	&	1.322	&	0.69		&	24656.11	&	25.17	\\
55203.59183	&	900		&	132	&	1.211	&	1.21		&	24658.71	&	28.74	\\
55204.63889	&	900		&	128	&	1.537	&	1.03		&	24686.85	&	26.46	\\
55254.54266	&	900		&	67	&	2.132	&	1.06		&	24674.11	&	24.01	\\
55255.52721	&	900		&	116	&	1.889	&	0.98		&	24700.37	&	37.09	\\
55256.53307	&	900		&	133	&	2.045	&	0.82		&	24692.35	&	26.00	\\
55258.52795	&	900		&	92	&	2.053	&	1.42		&	24642.86	&	28.91	\\
55262.54577	&	329		&	73	&	2.757	&	0.62		&	24608.74	&	52.33	\\
55263.54114$^\star$	&	546		&	103	&	2.294	&	0.80		&	24642.47	&	33.24	\\
55268.50341	&	900		&	97	&	2.110	&	--		&	24724.85	&	30.84	\\
55349.93534	&	800		&	60	&	1.889	&	--		&	24582.13	&	36.08	\\
55351.92877	&	800		&	98	&	1.908	&	1.22		&	24696.06	&	23.07	\\
55357.92508	&	800		&	95	&	1.718	&	--		&	24696.46	&	37.54	\\
55358.92916	&	800		&	84	&	1.635	&	1.65		&	24730.74	&	33.41	\\
55396.85224	&	1000	&	69	&	1.381	&	2.67	&		24733.71	&	34.97	\\
55399.84778	&	800		&	125	&	1.363	&	1.12		&	24725.86	&	31.32	\\
55400.91858	&	800		&	49	&	1.083	&	1.21		&	24732.45	&	27.47	\\
55401.88071	&	800		&	80	&	1.172	&	1.24		&	24729.91	&	22.26	\\
55402.87577	&	3000	&	46	&	1.147	&	3.45	&		24778.73	&	38.66	\\
55403.80577	&	800		&	47	&	1.640	&	1.23		&	24701.03	&	31.49	\\
55404.87298	&	800		&	143	&	1.171	&	0.49		&	24715.76	&	28.34	\\
55405.85977	&	800		&	118	&	1.212	&	0.44		&	24747.55	&	26.98	\\
55407.86952	&	800		&	101	&	1.155	&	1.08		&	24687.73	&	24.75	\\
55430.81255	&	800		&	129	&	1.137	&	0.96		&	24728.70	&	20.57	\\
55432.72912	&	800		&	71	&	1.611	&	1.40		&	24704.66	&	31.44	\\
55434.83699	&	800		&	134	&	1.066	&	0.65		&	24731.88	&	22.53	\\
55436.85315	&	800		&	156	&	1.048	&	--		&	24711.75	&	25.25	\\
55437.81918	&	600		&	128	&	1.082	&	0.40		&	24715.80	&	32.50	\\
55443.74436$^\star$	&	800		&	62	&	1.265	&	0.96		&	24667.47	&	27.30	\\
55491.59264	&	900		&	122	&	1.387	&	0.98		&	24625.13	&	36.85	\\
55492.72899	&	900		&	110	&	1.051	&	0.94		&	24556.40	&	38.67	\\
55493.79359	&	900		&	139	&	1.196	&	1.03		&	24708.20	&	27.49	\\
55494.84164	&	900		&	130	&	1.511	&	1.15		&	24672.08	&	38.83	\\
55497.80114	&	900		&	127	&	1.281	&	0.94		&	24675.84	&	20.47	\\
55521.65675	&	900		&	139	&	1.057	&	1.07		&	24693.18	&	22.35	\\
55523.64966	&	900		&	145	&	1.055	&	0.82		&	24677.85	&	26.95	\\
55524.57918	&	900		&	152	&	1.083	&	0.90		&	24675.97	&	22.39	\\
55525.65589	&	900		&	157	&	1.069	&	0.67		&	24725.37	&	25.32	\\
55526.64926	&	900		&	165	&	1.063	&	0.54		&	24714.46	&	27.10	\\
55527.63150	&	900		&	147	&	1.050	&	0.51		&	24708.28	&	17.16	\\
55583.63214	&	900		&	126	&	1.944	&	0.64		&	24661.91	&	34.04	\\
55584.59585	&	900		&	111	&	1.510	&	1.36		&	24737.89	&	26.59	\\
55585.61612	&	1000	&	141	&	1.785	&	0.70	&		24749.38	&	30.60	\\
55587.60350	&	1200	&	73	&	1.706	&	1.36	&		24700.16	&	19.31	\\
55588.62755	&	900		&	82	&	2.122	&	--		&	24707.08	&	35.28	\\
55907.60665$^\star$	&	900		&	159	&	1.064	&	0.61		&	24668.73	&	25.30	\\
55908.51258$^\star$	&	900		&	142	&	1.115	&	0.85		&	24679.38	&	17.49	\\
55909.64810$^\star$	&	900		&	139	&	1.170	&	0.94		&	24720.42	&	21.30	\\
55910.51765$^\star$	&	900		&	129	&	1.092	&	--		&	24686.16	&	22.72	\\
55911.64670$^\star$	&	900		&	147	&	1.184	&	1.19		&	24753.35	&	27.62	\\\noalign{\smallskip}
\hline                        
\noalign{\smallskip}
\end{longtable}
\tablefoottext{}{$^\star$: FEROS observations of HIP~11952 not present in the analysis of \citet{Setiawanetal2012}.}
\end{longtab}

\end{document}